\documentclass[aps,prb,twocolumn,amsmath,amssymb,showpacs,superscriptaddress]{revtex4-1}

\usepackage{graphicx}
\usepackage{amsmath,amssymb,amsfonts}
\usepackage{textcomp}
\usepackage{hyperref}
\usepackage{gensymb}
\usepackage{verbatim}
\usepackage{amsmath}
\hypersetup{breaklinks=true,colorlinks=true,urlcolor=black}
\usepackage{color}
\usepackage{soul}
\DeclareGraphicsExtensions{.jpg,.pdf,.png,.eps}

\begin{document}
\title{Anisotropic physical properties and pressure dependent magnetic ordering of CrAuTe$_{4}$}
\author{Na Hyun Jo, Udhara S. Kaluarachchi, Yun Wu, Daixiang Mou, Lunan Huang, Valentin Taufour, Adam Kaminski, Sergey L. Bud'ko and Paul C. Canfield}

\email[]{canfield@ameslab.gov}

\affiliation{Ames Laboratory and Department of Physics and Astronomy, Iowa State University, Ames, Iowa 50011, USA}

\date{\today}

\begin{abstract}
Systematic measurements of temperature dependent magnetization, resistivity and angle-resolved photoemission spectroscopy (ARPES) at ambient pressure as well as resistivity under pressures up to 5.25 GPa were conducted on single crystals of CrAuTe$_4$. 
Magnetization data suggest that magnetic moments are aligned antiferromagnetically along the crystallographic $c$-axis below $T_\textrm{N}$ = 255\,K. ARPES measurements show band reconstruction due to the magnetic ordering. Magnetoresistance data show clear anisotropy, and, at high fields,  quantum oscillations. The Neel temperature decreases monotonically under pressure, decreasing to $T_\textrm{N}$ = 236\,K at 5.22\,GPa. The pressure dependencies of (i) $T_\textrm{N}$, (ii) the residual resistivity ratio, and (iii) the size and power-law behavior of the low temperature magnetoresistance all show anomalies near 2\,GPa suggesting that there may be a phase transition (structural, magnetic, and/or electronic) induced by pressure. For pressures higher than 2\,GPa a significantly different quantum oscillation frequency emerges, consistent with a pressure induced change in the electronic states.
  
\end{abstract}

\maketitle 

\section{Introduction}
Transition metal based antiferromagnetism (AFM) can be considered to be a necessary, but clearly not sufficient, ingredient for high temperature superconductivity in both the CuO- and Fe-based superconductors.  When long-range AFM can be suppressed by either substitution or pressure in these systems superconductivity emerges, often with remarkably high transition temperatures and upper critical fields.  One proposed route to the discovery of new high-Tc superconductors is through the study of other transition metal based antiferromagnets, specifically their response to pressure.\cite{CanfieldB2016} Many studies have been done on Cr based AFM materials not only at ambient pressure but also under pressure. Pure elemental Cr showed exponential decrease of $T_\textrm{N}$ as pressure was applied.\cite{McWhan1967,Fawcett1988,Fawcett1994} V doped Cr also showed a substantial decrease in $T_\textrm{N}$,\cite{Yeh2002} and several studies examined the effects of pressure on V doped Cr to see quantum criticality.\cite{Lee2004,Feng2007,Jaramillo03082010} More recently, pressure studies on CrAs, which has $T_N$ = 265\,K at ambient pressure, demonstrate complete suppression  of AFM, and even a superconducting transition around 2\,K at the critical pressure of 8\,kbar.\cite{Wu2014} All of these pressure studies are important, because new physics around the quantum critical point can be studied.\cite{Coleman2005} 

CrAuTe$_4$ was reported to crystallize in a monoclinic space group ($P2/m$) with lattice parameters $a$\,=\,5.4774(7)\,$\AA$, $b$\,=\,4.0169(6)\,$\AA$, $c$\,=\,7.3692(13)\,$\AA$ and $\beta$\,=\,90.604(10)\,$^{\circ}$.\cite{Reynolds2004} It has only one Cr site with monoclinic centrosymmetric point symmetry (2/m), and it reported AFM ordering temperature was below 255\,K.\cite{Reynolds2004} Previous studies were done on polycrystalline samples and only at ambient pressure.

In this paper, we report the physical properties of single crystal CrAuTe$_4$ at ambient pressure as well as temperature and field dependent resistivity for pressures up to 5.22\,GPa. We find that CrAuTe$_4$ orders antiferromagnetically with $T_\textrm{N}$ = 255\,K at ambient pressure, with the ordered moments along the crystallographic $c$-axis.  Angle resolved photoemission spectroscopy (ARPES) data show that there is a clear reconstruction of the Fermi Surface associated with the magentic ordering and pressure dependent resistivity measurements show that the ordering temperature can be suppressed to 236\,K by 5.22\,GPa.  In addition to the suppression of the antiferromagnetic ordering temperature we are able to infer that for $P$\,$\sim$\,2\,GPa there is an additional, pressure induced phase transition that leads to clear changes in the temperature and field dependent resistivity.  

\section{Experiment}
Single crystals of CrAuTe$_{4}$ were grown by adding small amount of Cr to the low temperature Au-Te eutectic.\cite{ASM2000} High purity elemental Cr, Au and Te were put into alumina crucible with initial stoichiometry, Cr$_{2}$Au$_{30}$Te$_{68}$, and sealed in amorphous silica tube.\cite{Canfield2016,Canfield1992} The ampules were heated up to 900\,\celsius\, within 5 hours, kept it for 3 hours, rapidly cooled to 625\,\celsius\, and then slowly cooled down to 500\,\celsius\, over 75 hours, and finally decanted using a centrifuge.\cite{Canfield1992} We always obtained both bladelike CrAuTe$_{4}$ with typical dimensions of $5 \times 0.2 \times 0.1$ mm and platelike Cr$_{5}$Te$_{8}$ at the same time. Identifying the two compounds was possible given their different morphologies as well as their different magnetic properties. Since Cr$_{5}$Te$_{8}$ is a ferromagnet,\cite{Ohta1993} even a small amount of Cr$_{5}$Te$_{8}$ could be detected by measuring magnetic properties. 

\begin{figure}
	\includegraphics[scale=1]{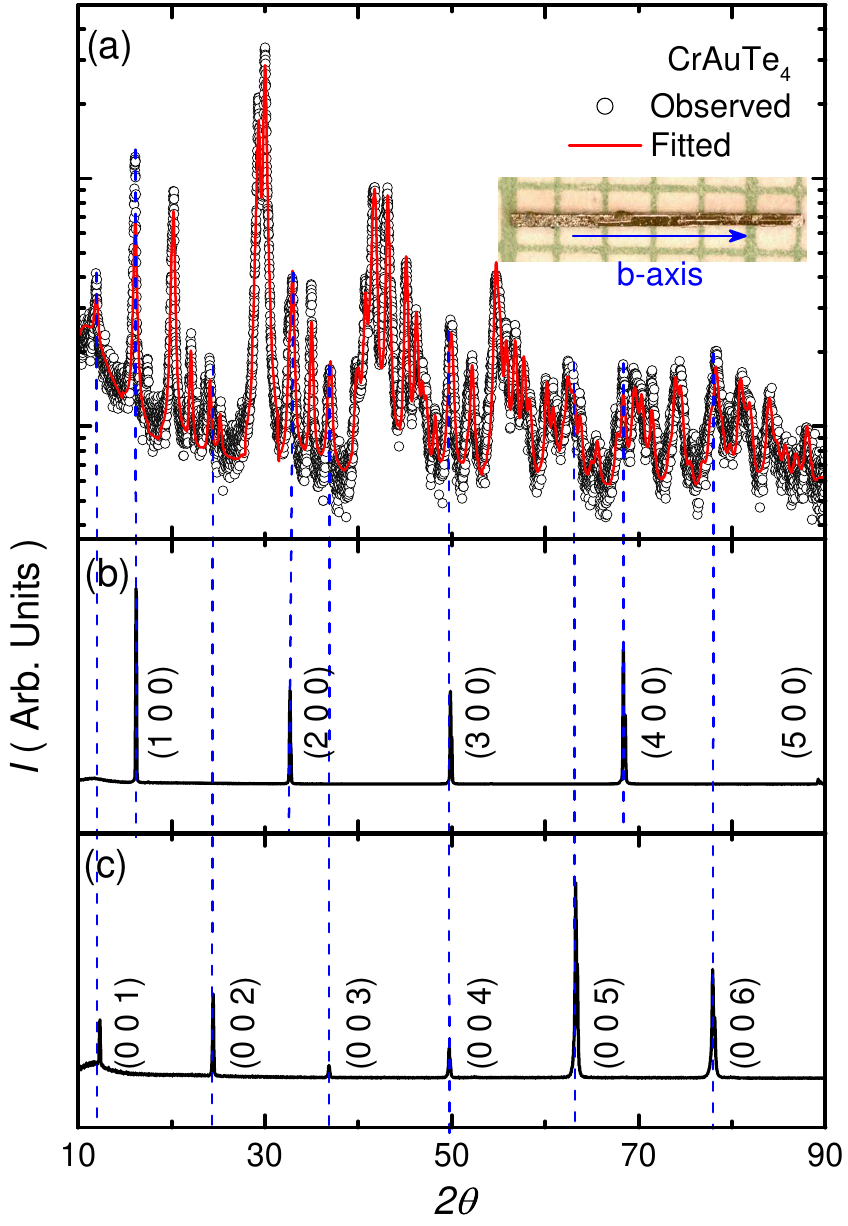}%
	\caption{(color online) (a) Observed and fitted powder x-ray diffraction (XRD) pattern of CrAuTe$_{4}$. Inset picture is typical CrAuTe$_{4}$ single crystal with the $b$-axis as the longest edge. (b) Single crystal XRD along the $a$-axis. (c) Single crystal XRD along the $c$-axis.
		\label{xrd}}
\end{figure}

A Rigaku MiniFlex diffractometer (Cu $K_{\alpha1,2}$ radiation) was used for acquiring x-ray diffraction (XRD) pattern at room temperature. The powder pattern was fitted with Le Bail refinement (Fig.\,\ref{xrd}\,(a)). Lattice parameters obtained from this refinement were $a$=5.49$\AA$, $b$=4.02$\AA$, $c$=7.37$\AA$, and the angles $\alpha$=90$\degree$, $\beta$=90.64$\degree$, $\gamma$=90$\degree$ in agreement with reported values.\cite{Reynolds2004}.

The orientation of the crystal was determined by single crystal XRD.\cite{Jesche2016} When the largest surface of the crystal was exposed to XRD, only (h 0 0) peaks, where h is integer number, were detected (Fig.\,\ref{xrd}\,(b)). After confirming the $a$-axis direction, we rotated the sample almost 90$\degree$ ($\beta$=90.64$\degree$) and measured XRD again to confirm the $c$-axis direction; in this case, only (0 0 l) peaks were detected (Fig.\,\ref{xrd}\,(c)). By elimination, the long axis of the crystal is the crystallographic $b$-axis. 

Magnetic measurements were performed in a Quantum Design, Magnetic Property Measurement System (MPMS), SQUID magnetometer. The sample was mounted between two long straws for $H \parallel$ b. Two thin films, which have small diamagnetic signals, were used to hold the sample for $H \parallel$ a and $H \parallel$ c magnetization measurements. Temperature and field dependent electrical transport measurement were carried out in a Quantum Design Physical Property Measurement System (PPMS) for $1.8 \,\textrm K \leq T \leq 305 \,\textrm K$ and $\mid H \mid \leq 90 \,\textrm {kOe}$. Samples for transport measurement were prepared by attaching four Au wires (12.7\,$\mu$m diameter) using spot welding. The contact resistance values were all around 1 $\ohm$. 

ARPES data were acquired using a tunable, vacuum ultraviolet laser-based ARPES \cite{Jiang2014}. The samples were cleaved \textit{in situ} at 40\,K at a base pressure lower than 8 $\times$ 10$^{-11}$ Torr.  Momentum and energy resolution were set at $\sim$0.005 \AA$^{-1}$ and 2\,meV, respectively. 

Two types of pressure cells were used for this experiment. A Be-Cu/Ni-Cr-Al hybrid piston-cylinder cell (PCC), similar to that used in Ref.\,\onlinecite{Budko1984}, was used for pressures up to  2.1\,GPa. For this pressure cell, a 4:6 mixture of light mineral oil: $n$-pentane\,\cite{Budko1984,Kim2011PRB} was used as a pressure medium. It solidifies at $\sim$3-4\,GPa at room temperature\cite{Kim2011PRB,Torikachvili2015}. For higher pressure, a modified Bridgman anvil cell (mBAC)\cite{Colombier2007,Torikachvili2015} was used with a 1:1 mixture of $n$-pentane:iso-pentane as a pressure medium which is considered as a good hydrostatic medium\cite{Tateiwa2010,Klotz2009,Kim2011PRB,Torikachvili2015}. The solidification of this medium occurs around $\sim$6-7\,GPa at room temperature\cite{Piermarini1973,Klotz2009,Kim2011PRB,Torikachvili2015}. For both cells, the pressure was determined by the superconducting transition temperature of Pb\cite{Bireckoven1988} measured by the resistivity.  

\section{Results and Analysis}

\subsection{\texorpdfstring{Magnetization}{space}}

\begin{figure}
	\includegraphics[scale=1]{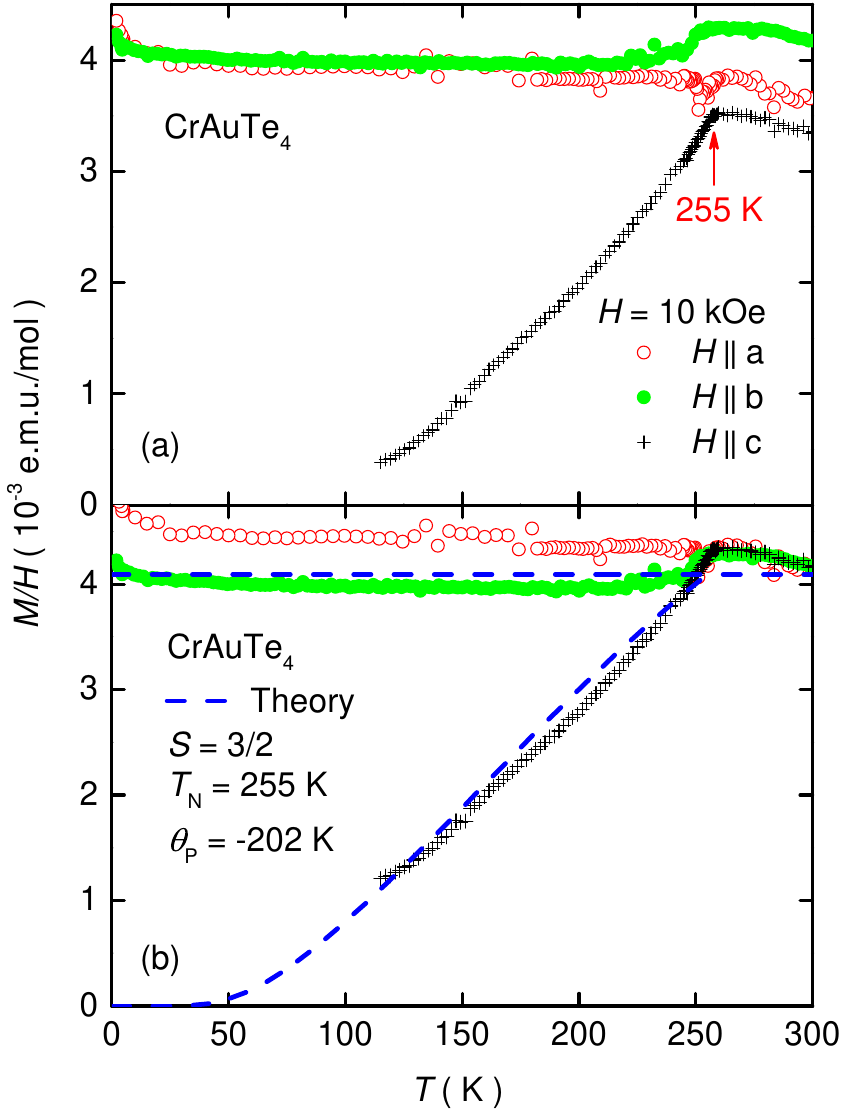}%
	\caption{(color online) (a) Anisotropic magnetization data along three different directions. Black cross represents the magnetization when $H \parallel$ c, green filled dot corresponds to $H \parallel$ b magnetization, and red open dot is the magnetiztion with $H \parallel$ a. (b) anisotropic magnetization data with $H \parallel$ a and $H \parallel$ c data shifted additively to compensate for small addenda contribution (see text). Dashed lines are from a MFT result (see text). 
		\label{Magnetic}}
\end{figure}

 In Fig.\,\ref{Magnetic}\,(a), magnetic susceptibility as a function of temperature, $M(T)/H$, measured in $H = 10$\,kOe shows the AFM transition just above 250\,K. Because we mounted sample with diamagnetic thin film(press-and-seal wrap, the Glad Products Company), the magnetic susceptibility data for $H \parallel$ a and $H \parallel$ c have a diamagnetic background, which is a likely cause for the differences above $T_\textrm N$. In addition, no anisotropy is expected in the paramagnetic state of Cr$^{3+}$ in octahedron environment with S = 3/2 considering the crystalline electric field effect. Therefore, the H\,$\parallel$\,a and H\,$\parallel$\,c susceptibilities have been additively shifted to overlap the high temperature H\,$\parallel$\,b data in Fig.\,\ref{Magnetic}\,(b). 
 
 Magnetic susceptibility data below 250 K (in magnetically ordered state) are highly anisotropic between $H \parallel$\,c and $H \perp$\,c with an anisotropy of $\sim$\,10 at around 100\,K. We were not able to take the data below 100\,K for $H \parallel$\,c, because magnetization of our single crystal dropped below $\sim10^{-6} \,e.m.u.$ which is below the MPMS resolution. Simple visual inspection of the anisotropic data suggests that the ordered Cr moments are along the $c$-axis.

More quantitatively, the anisotropic, temperature dependent magnetic susceptibility of AFM systems can be analyzed within the context of mean field theory (MFT).\cite{Johnston2012, Johnston2015} The magnetic susceptibility with H perpendicular to the ordered magnetic moment below $T_\textrm{N}$ can be described by MFT. 
\begin{equation}
\frac { { \chi  }_{ \bot  }(T\le { T }_\textrm{ N }){ T }_\textrm{ N } }{ C } =\frac { 1 }{ 1-\theta_\textrm{P}/T_\textrm{N} } 
\end{equation} 
where theta $\theta_\textrm{P}$ is the paramagnetic Weiss temperature and, 
\begin{equation}
	C=2[S(S+1)]^{1/2}
\end{equation}

Based on previous magnetization data on polycrystalline CrAuTe$_4$,\cite{Reynolds2004} Cr is known to be trivalent. Cr$^{3+}$ has $S$ = 3/2 and the corresponding $C$ was calculated by using $C$ = 2[$S$($S$+1)]$^{1/2}$. We used $T_N$ = 255 K which is determined from both our magnetization and resistivity data (see below). The only free parameter that has to be determined is $\theta_\textrm P$. By matching the theoretical prediction and the magnetic susceptibility data for $H\,\parallel$\,b, which has almost no background, we find  $\theta_\textrm P$ = -202\,K. 

From this value, the magnetic susceptibility with field along the easy axis, $H\parallel c$, of collinear antiferromagnet below $T_N$ can be calculated without any free parameters: 
\begin{equation}
	\frac { { \chi  }_{ \parallel  }(T) }{ { \chi  }({ T }_\textrm{ N }) } =\frac { 1-\theta_\textrm{P}/T_\textrm{N} }{ { \tau  }^{ * }-\theta_\textrm{P}/T_\textrm{N} }  
\end{equation}
where 
\begin{equation}
	 { \tau  }^{ * }(t)=\frac { (S+1)t }{ 3{ B }_{ S }^{ ' }({ y }_{ 0 }) }   
\end{equation}

In the function $\tau^{*}(t)$, $t$ and $ { B }_{ S }^{ ' }({ y }_{ 0 })$ are 
\begin{equation}
	t\equiv \frac { T }{ { T }_\textrm{ N } },\quad { B }_{ S }^{ ' }({ y }_{ 0 })=[d{ B }_{ S }(y)/dy]{ | }_{ y={ y }_{ 0 } } 
\end{equation}
where $B_{S}(y)$ is unconventional Brillouin function which is defined as 
\begin{equation}
	{ B }_{ S }(y)=\frac { 1 }{ 2S } \left\{ (2S+1)coth[(2S+1)\frac { y }{ 2 } ]-coth\left( \frac { y }{ 2 }  \right)  \right\} 
\end{equation}
and ${ y }_{ 0 }=\frac { 3\overline { { \mu  }_{ 0 } }  }{ (S+1)t }$. $\overline { { \mu  }_{ 0 } }$ can be calculated numerically by graphical method from $\overline { { \mu  }_{ 0 } }=B_{S}(y_{0})$. The calculated behavior, based on MFT, is shown in Fig.\,\ref{Magnetic}\,(b) with dashed blue lines. The MFT calculated anisotropic susceptibility is consistent with the experimental data.

\subsection{\texorpdfstring{Resistivity}{space}}

\begin{figure}
	\includegraphics[scale=1]{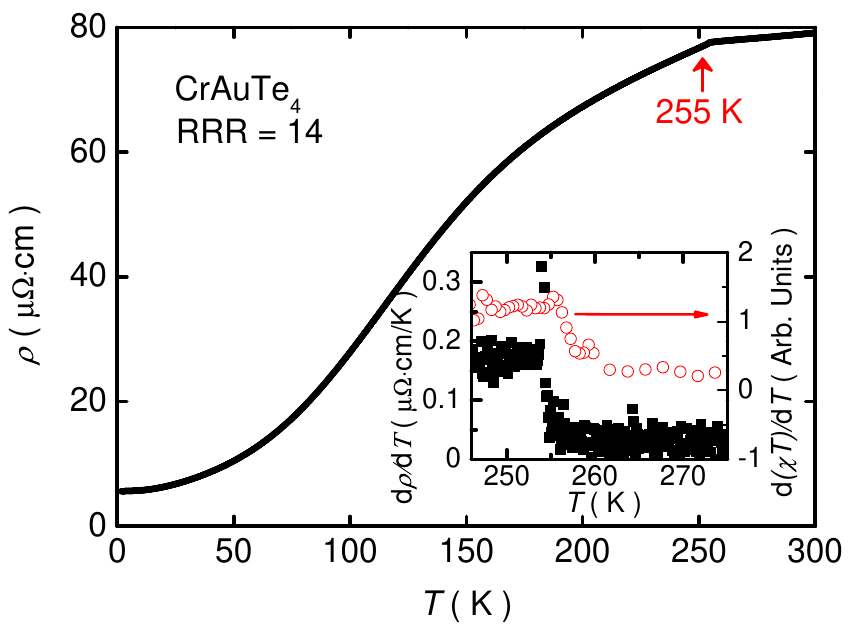}%
	\caption{(color online) Zero-field, temperature dependent resistivity $\rho (T)$ of CrAuTe$_{4}$. Filled black squares show derivative of $\rho (T)$ with respect to temperature and open red circles represent the $d(\chi T)/dT$. Both shows a clear peak at 255 K. 
		\label{RT}}
\end{figure}

The electrical resistivity, $\rho (T)$, of CrAuTe$_{4}$ was measured from 1.8\,K to 305\,K. A kink, where an abrupt slope change occurs, is found at $T_\textrm N$, in Fig.\,\ref{RT}. The residual resistivity ratio (RRR) determined by $\rho (300\,\textrm K)/\rho (2\,\textrm K)$ of the sample is 13 with $\rho_{0} = 5.566\,\mu\ohm\cdot \textrm {cm}$. The low residual resistivity and high RRR demonstrate the good quality of the single crystal. In order to evaluate the transition temperature from different measurements, we used Fisher's arguments of singular behavior near an AFM transition transition.\cite{Fisher1962, Fisher1968} It is expected that the anomaly shown in specific heat due to AFM material is similar to anomaly in $d(\chi T)/dT$,\cite{Fisher1962} and that the resistivity anomaly, $d\rho/dT$, due to short-range fluctuation near transition temperature exhibits similar feature as specific heat.\cite{Fisher1968} We calculated $d(\chi T)/dT$ and $d\rho/dT$ as shown in Fig.\,\ref{RT} inset. Both $d(\chi T)/dT$ and $d\rho/dT$ show step like features just above and just below 255\,K (respectively), giving $T_\textrm N$\,=\,255\,K\,$\pm$\,1\,K. 

\begin{figure}
	\includegraphics[scale=1]{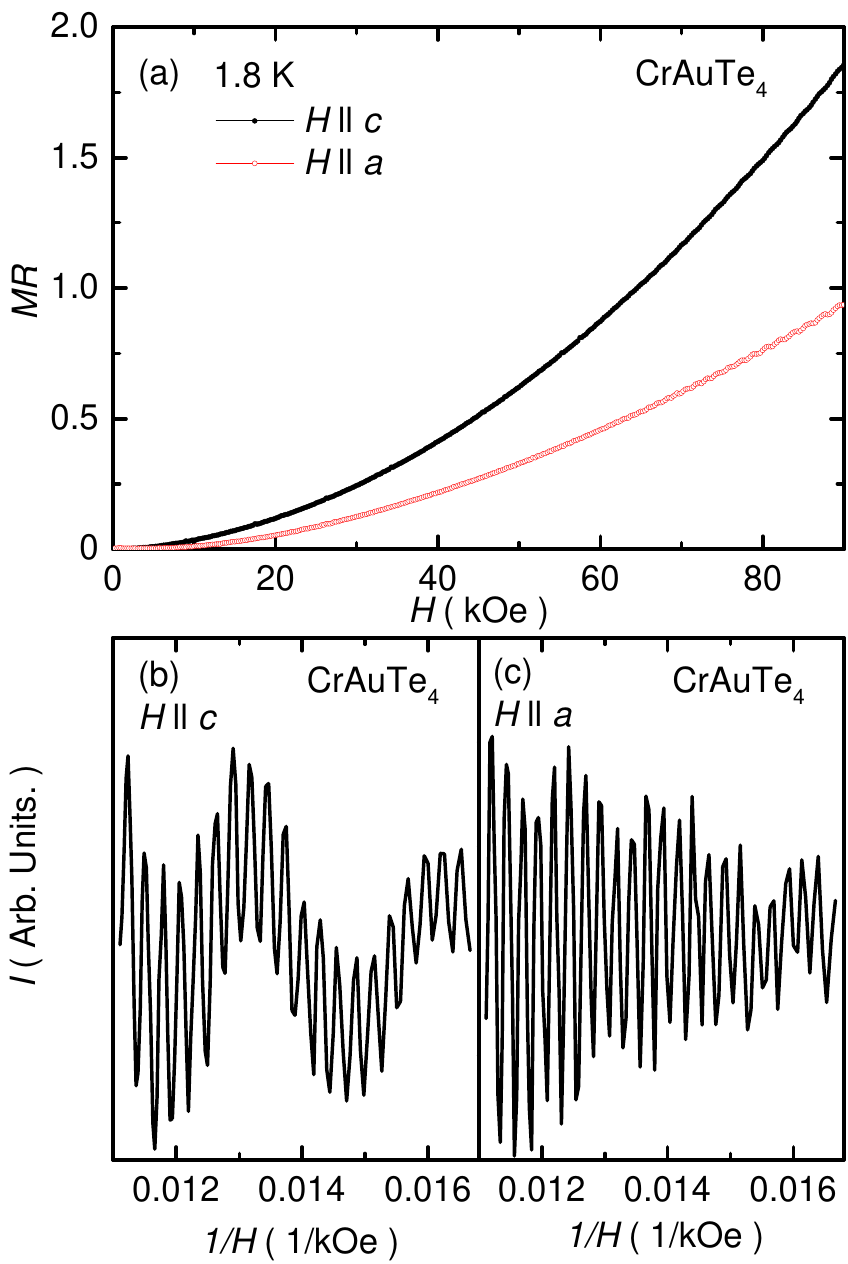}%
	\caption{(color online) (a) Magnetoresistance (MR = $\rho(H)$/$\rho(H=0)$)  of CrAuTe$_{4}$ at 1.8\,K for two different field directions. (b) and (c), Shubnikov-de Haas oscillations after subtracting a second order polynomial background. (b) magnetic field along the $c$ direction. (c) magnetic field along the $a$ direction.
		\label{MR}}
\end{figure}

The field dependent resistivity was measured up to 90\,kOe at 1.8\,K along two different directions (Fig.\,\ref{MR}). When the field direction is H\,$\parallel$\,c, the magnetoresistance (MR = ($\rho(H)$-$\rho(H=0)$)/$\rho(H=0)$) for a fixed temperature is 1.85 at 90\,kOe. On the other hand, MR is 0.94 at 90\,kOe for H\,$\parallel$\,a. In the high field regime, Shubnikov de Haas (SdH) oscillations become increasingly apparent. We subtract the background by using second order polynomial function and plot the obtained oscillatory part of MR as a function of 1/H in Fig.\,\ref{MR}\,(b) and (c). Two distinct frequencies are found for H\,$\parallel$\,c. On the other hand, a beat-like feature, due to two adjacent frequencies, is detected for H\,$\parallel$\,a.   

\begin{figure}
	\includegraphics[scale=1]{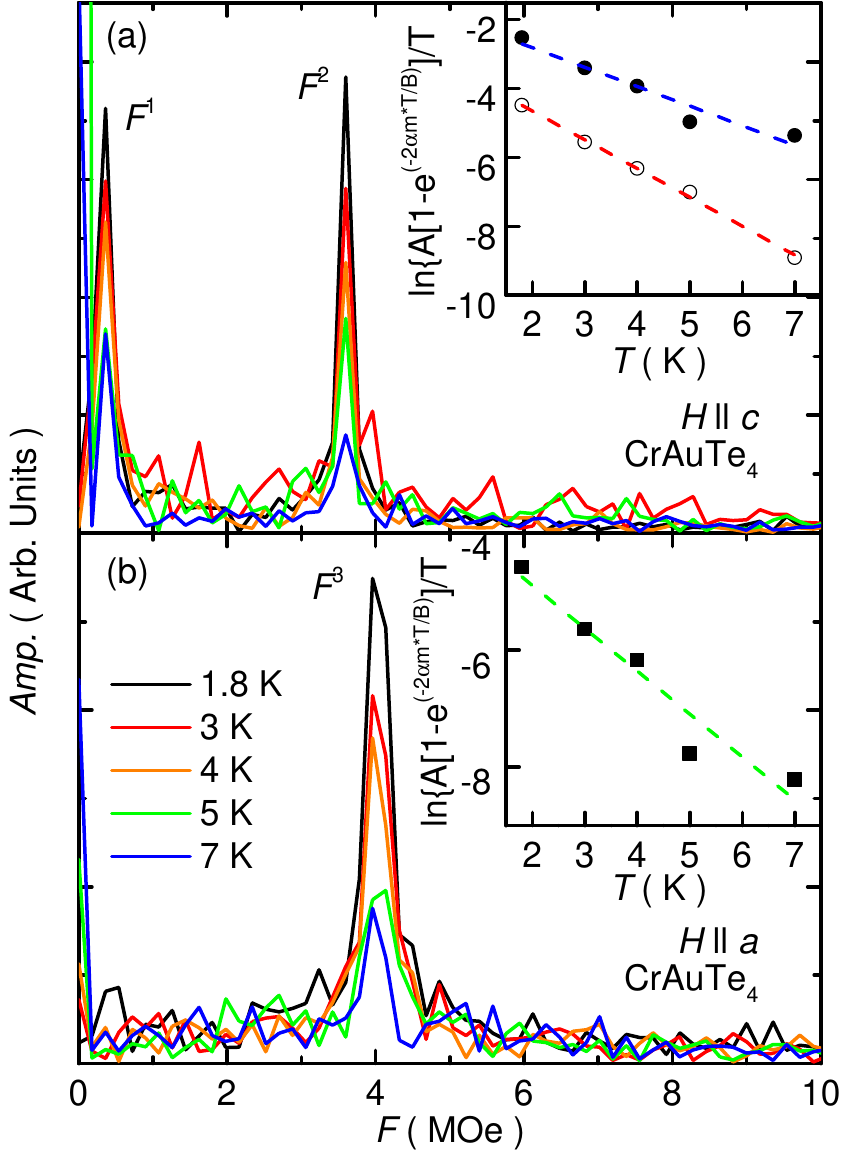}%
	\caption{(color online) (a) Fast Fourier Transform (FFT) analysis of quantum oscillation data for $T$\,=\,1.8\,K, 3\,K, 4\,K, 5\,K and 7\,K when magnetic field is applied along the $c$-axis. Inset is the mass plot. The closed circle is the data from $F^1$ and open square is the data from $F^{2}$. (b) Same as (a) except the magnetic field direction on the crystal was along the $a$-axis.    
		\label{SdH}}
\end{figure}

We use a Fast Fourier transform (FFT) algorithm to compare the quantum oscillation for the two directions of applied field. (Fig.\,\ref{SdH}) The FFT result for H $\parallel$ c shows two sharp peak; $F^{1}\,=\,0.36$\,MOe and $F^{2}\,=\,3.6$\,MOe. At first glance only one frequecy is found from H\,$\parallel$\,a: $F^{3}\,=\,3.9$\,MOe, but, if we look at the FFT data for H\,$\parallel$\,a more carefully, we can see that the peak is broader that those found for H\,$\parallel$\,c. This may indicate that it indeed has two adjacent frequencies associated with the beat like feature seen in Fig.\,\ref{MR}\,(c). From the Onsager relation, $F^{i} = \frac{\hbar c}{2 \pi e}S^{i}$, which states direct proportionality between frequency and extremal areas, three corresponding extremal area of the Fermi surface were calculated: $S_{F^1} = 0.00344\,\AA^{-2}$, $S_{F^2} = 0.0344\,\AA^{-2}$ and $S_{F^3} = 0.0378\,\AA^{-2}$.

MR was also measured at 5 different temperatures; 1.8\,K, 3\,K, 4\,K, 5\,K, and 7\,K, and the FFT is done on each MR data set (Fig.\,\ref{SdH}) with same data processing. The amplitude of quantum oscillations decreases as temperature increases which indicates a phase smearing at finite temperature. This can be explained by Lifshitz-Kosevich formula. \cite{Shoenberg1984} The formula states that finite temperature, finite electron relaxation time and electron spin can introduce the phase smearing 
    \begin{equation}
    { A }_{ t }\propto { B }^{ 1/2 }{ \left| \frac { { \partial  }^{ 2 }S_{ t } }{ \partial { k }_{ H }^{ 2 } }  \right|  }^{ -1/2 }{ R }_{ T }{ R }_{ D }{ R }_{ S }
    \end{equation} 
where the factor $R_{T}$ is related to thermal damping, $R_{D}$ is related to impurities and $R_{S}$ is related to spin. Specifically, the thermal damping part $R_{T}$ is defined as  
 \begin{equation}
 { R }_{ T }=\frac { \alpha { m }^{ * }T/B }{ sinh(\alpha { m }^{ * }T/B) } 
 \end{equation} 
where $\alpha=2\pi^{2} c k_{B}/e \hbar $. Thus, we can obtain the effective mass $m^{*}$ by linear fitting using the mass plot in Fig.\,\ref{SdH} inset. The calculated effective mass for $F^{1}$ is 0.19 $m_{e}$, $F^{2}$ is 0.25 $m_{e}$ and $F^{3}$ is 0.25 $m_{e}$. The effective masses for $F^{2}$ and $F^{3}$ are same, and their frequencies are also almost same as well. These features in $F^{2}$ and $F^{3}$ might indicate that these frequencies are associated with a spherical sheet of the Fermi surface whereas $F^{1}$ is not. 

\subsection{\texorpdfstring{ARPES}{space}}
The clear signatures of $T_\textrm{N}$ in magnetization and resistivity are accompanied with changes in the Fermi surface as well. Whereas the SdH oscillations can provide some information about the low temperature, extremal orbit dimension, a $T_\textrm{N}$ value of 255 K makes quantum oscillation measurements for $T$\,$>$\,$T_\textrm{N}$ highly unlikely.  Fortunately, ARPES measurements can probe the Fermi surface both above and below $T_\textrm{N}$. 

\begin{figure}
	\includegraphics[scale=0.5]{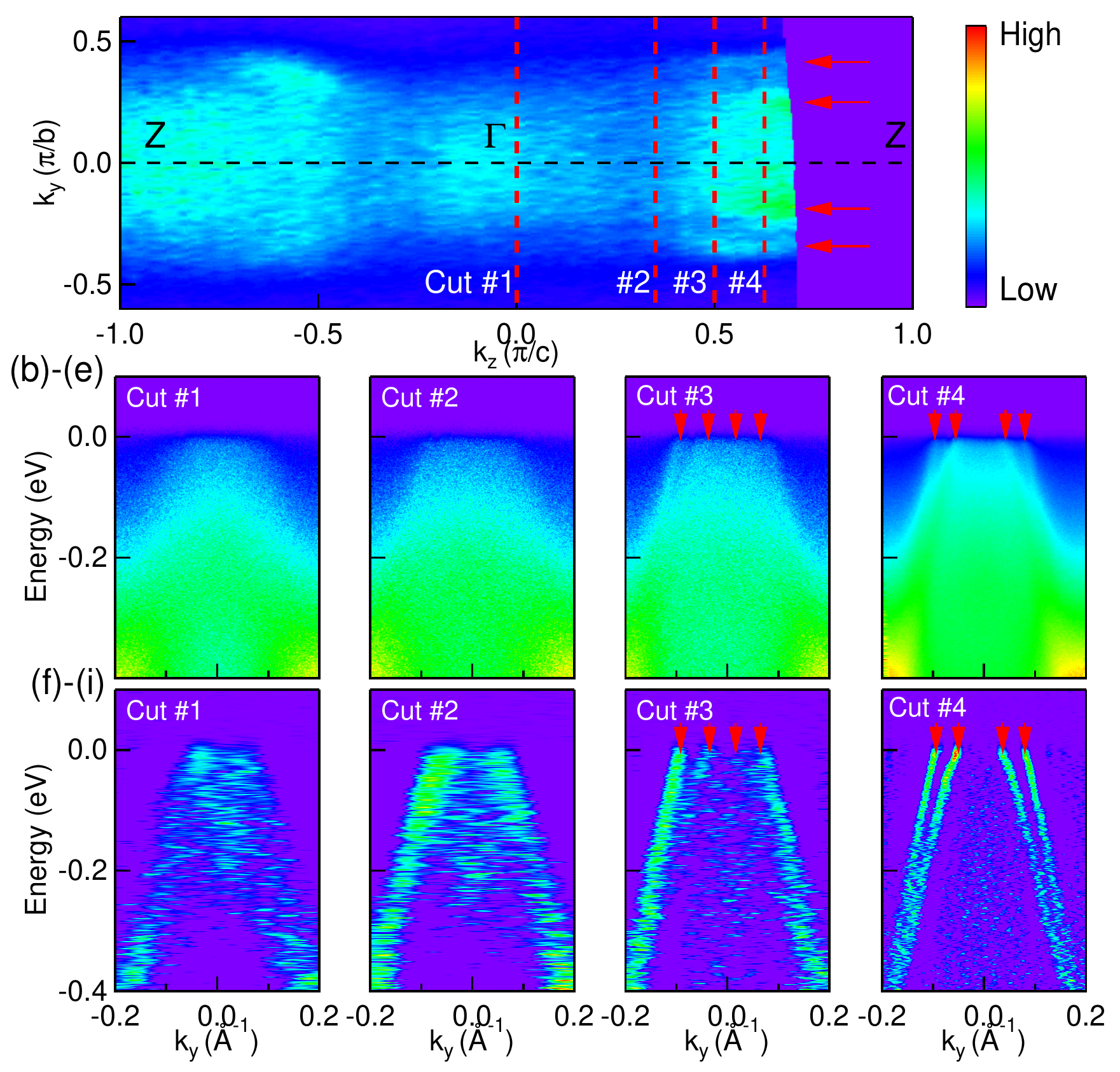}%
	\caption{(color online) Fermi surface plot and band dispersion measured using 6.70 eV photon energy. 
		(a) Plot of ARPES intensity integrated within 10 meV about the chemical potential measured at $T=$40 K. High intensity areas correspond to location of Fermi sheets and/or proximity of the bands to the chemical potential. The red arrows indicate the location of the Fermi momenta found in figures (e) and (i) below. 
		(b)-(e) ARPES intensity along cuts 1-4 measured at $T= 40$. 
		(f)-(i) The intensity plots of the second derivatives of data in (b), (c), (d) and (e). The red arrows in (d),(e), (h) and (i) mark the the locations of Fermi momenta. 
		\label{Arpes1}}
\end{figure}

In Fig.\,\ref{Arpes1}, we show the Fermi surface plot and band dispersion measured using 6.70 eV photon energy. Fig.\,\ref{Arpes1}\,(a) shows the ARPES intensity in the first Brillouin zone, integrated within 10 meV about the chemical potential. The red arrows point to the location of two Fermi sheets. In Figs.\,\ref{Arpes1}\,(b)-(e), we show the band dispersion along cuts 1-4 (marked in panel (a)) measured at $T=$ 40\,K. As we move away from $\Gamma$ to $Z$ (i.e. from cut 1 to cut 4 in Fig.\,\ref{Arpes1}), two hole pockets emerge as indicated by the red arrows, which can be better seen in the second derivative plots of the ARPES intensity shown in Figs.\,\ref{Arpes1}\,(f)-(i).

\begin{figure}
	\includegraphics[scale=0.8]{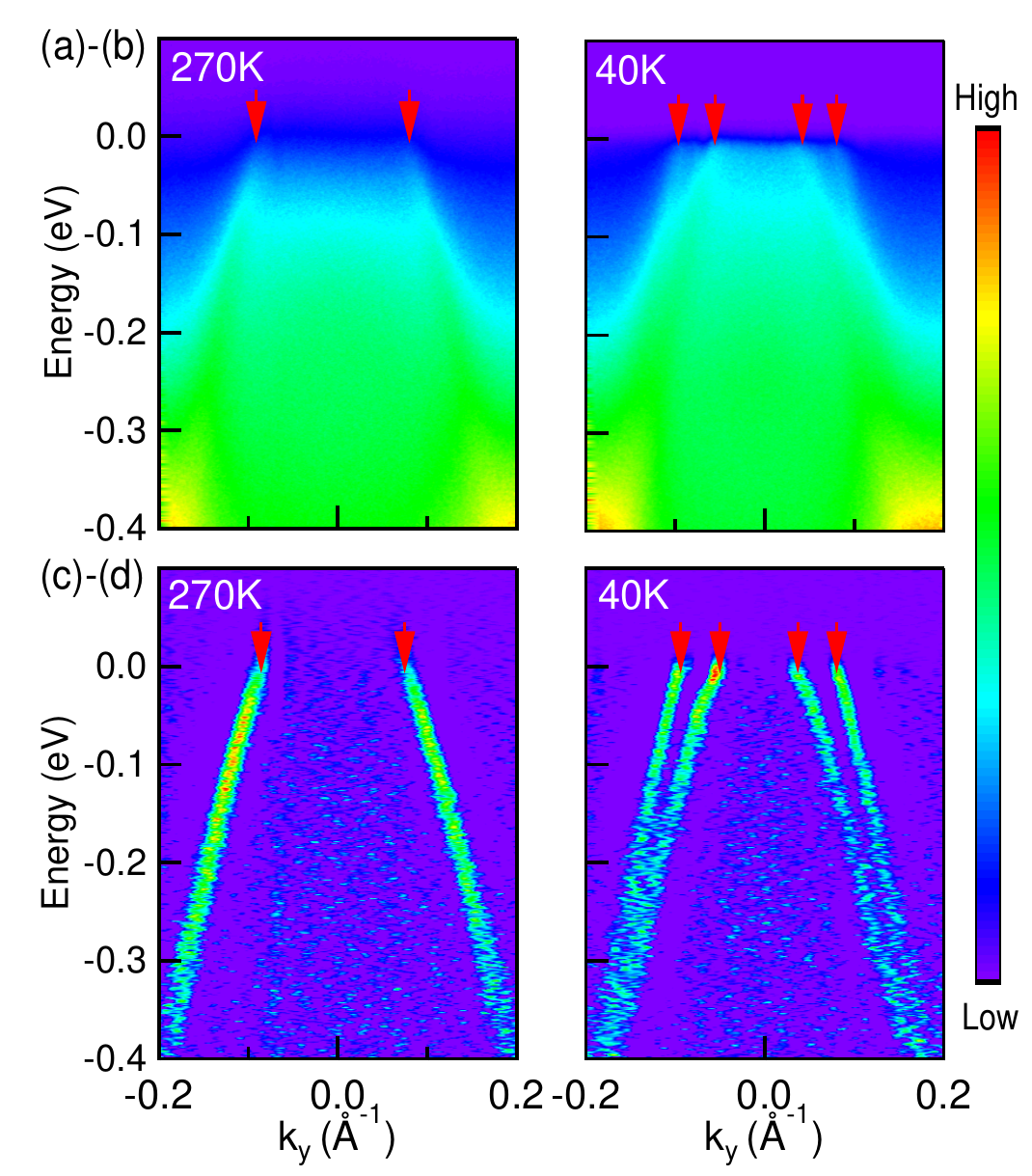}%
	\caption{(color online) Band dispersion measured using 6.70 eV photon energy. 
		(a)-(b) ARPES intensity along cut 1 measured at $T= $270 and 40\,K, respectively. 
		(c)-(d) The intensity plots of the second derivatives of data in (a) and (b). The red arrows mark the Fermi crossings. 
		\label{Arpes2}}
\end{figure}

In Figs.\,\ref{Arpes2}\,(a)-(b), we show the band dispersion along cut 4 in Fig.\,\ref{Arpes1}\,(a) measured at two temperatures: 270 and 40\,K (above and below the AFM transition temperature, respectively). We can clearly see that above the AFM transition temperature of 255\,K (in the paramagnetic state), there is only a single hole pocket (marked by red arrows). When the sample temperature is decreased to 40\,K, another hole pocket emerges in the center. To better illustrate the band dispersion, we calculate and plot second derivative of the ARPES intensity in Figs.\,\ref{Arpes2}\,(c)-(d). 

Diameters of two pockets that we marked on Fig.\,\ref{Arpes2} at 40 K are $0.1708\,\AA^{-1}$ and $0.0815\,\AA^{-1}$. Assuming circular orbits, the calculated surface areas are $0.0229\,\AA^{-2}$ and $0.005\,\AA^{-2}$. Former one may corresponds to $F^{2}$ and $F^{3}$. However, the latter one is not well matching. The differences between SdH data and ARPES data are due to limited access to full Brillouin zone along the k$_x$ direction. Thus, the values that come from ARPES data often are close to, but different from the extremal value inferred from SdH oscillations. 

The Fermi surface topology change is due to the magnetic transition from paramagnet to AFM. In the AFM state, the unit cell size can change, and it can affect the Brillouin zone in k-space. Thus, it introduces the band folding in the system.\cite{liu2010,Kondo2010} This is the reason why two bands were observed below $T_\textrm{N}$, although only one band was observed above the transition temperature. 

\subsection{\texorpdfstring{Resistivity under pressure}{space}}

The temperature dependent electrical resistivity of CrAuTe$_4$ was measured up to 5.22 GPa.  This was accomplished by using two different pressure cells (see experimental methods section) each with its own sample.  Both samples had RRR values similar to that shown in Fig.\,\ref{RT}: RRR = 14 for the sample used in the PCC and RRR = 13 for the sample used in the mBAC.

\begin{figure}
	\includegraphics[scale=1]{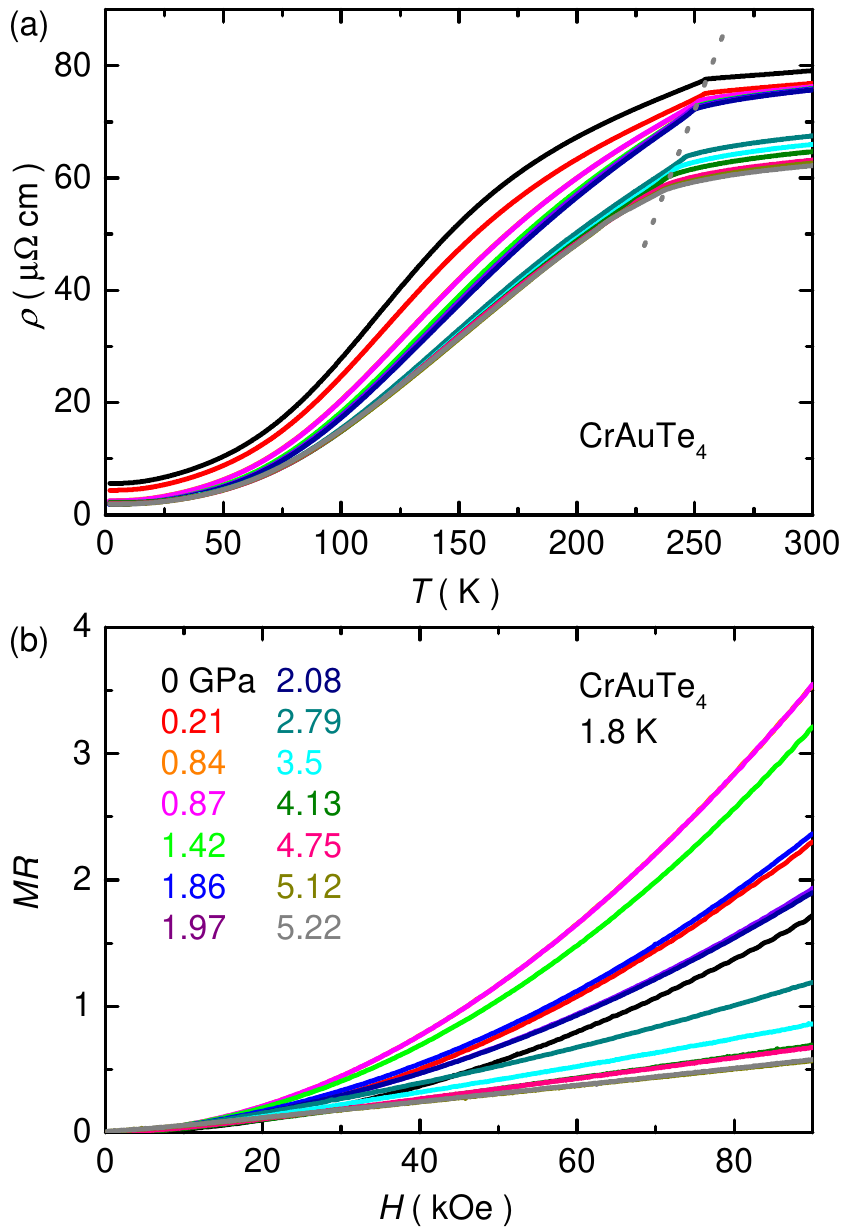}%
	\caption{(color online) Pressure studies on CrAuTe$_{4}$. (a) Temperature dependent resistivity at 0, 0.21, 0.84, 0.87, 1.42, 1.86, 1.97, 2.08, 2.79, 3.50, 4.13, 4.75, 5.12 and 5.22\,GPa. (b) MR as a function of field at 1.8\,K with same pressures as (a). 
		\label{Pressure}}
\end{figure}

\begin{figure}
	\includegraphics[scale=1]{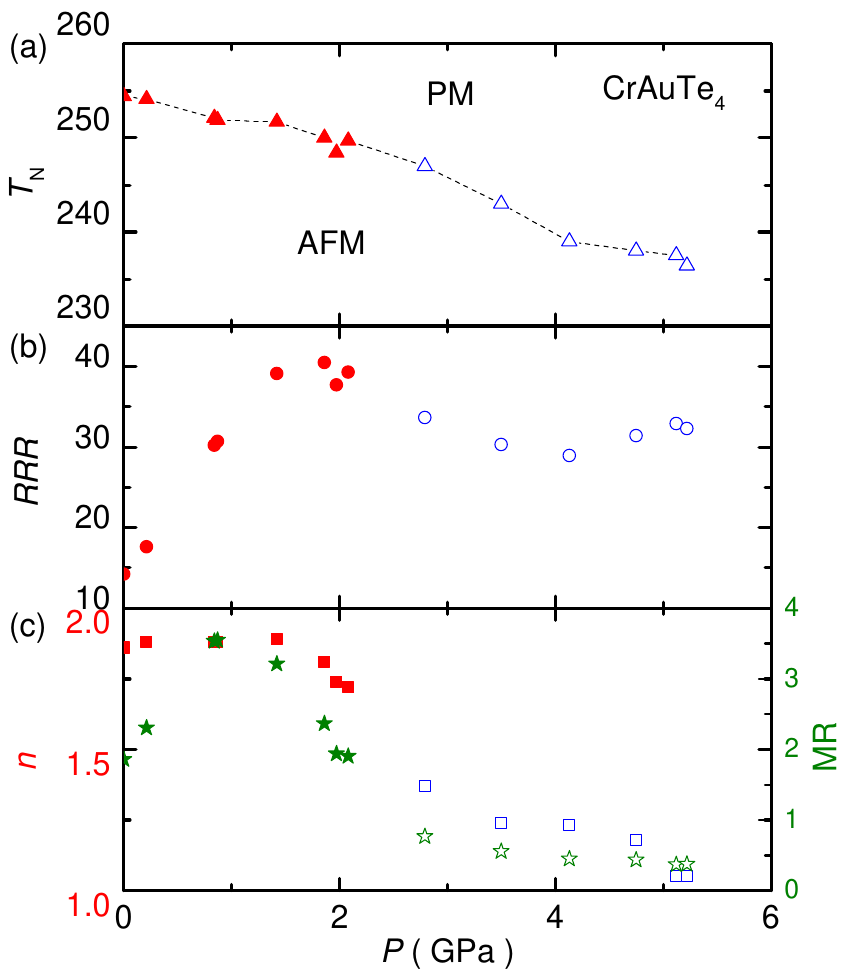}%
	\caption{(color online) Pressure dependence of transport properties: (a) transition temperature (filled red triangle data was measured in piston-cylinder cell, open blue triangle data was measured in Bridgeman cell) (b) RRR (filled red circle data was measured in piston-cylinder cell, open blue circle data was measured in Bridgeman cell) (c) (left) square stands for exponent of H (filled red square data from piston-cylinder cell, open blue square data from Bridgeman cell), (right) green stars are MR at 90 kOe (filled data from piston-cylinder cell, open data from Bridgeman cell).  
		\label{Pressure_anal}}
\end{figure}

The temperature dependence of the resistivity at various pressures is shown in Fig.\,\ref{Pressure}\,(a). The data from each of the two samples was normalized to the ambient pressure resistivity value from Fig.\,\ref{RT}. The antiferromagnetic transition temperature decreases as we apply pressure.  $T_\textrm{N} = 254$\,K inferred from the ambient pressure decreases monotonically to $T_\textrm{N} = 236$\,K at 5.22\,GPa (see Fig.\,\ref{Pressure_anal}\,(a)). The rate is -3.4\,K/GPa when we assume a linear change. In fact, though, there are two breaks in slope of $T_\textrm{N} (P)$, one at roughly 2 GPa and a second at roughly 4 GPa.  Whereas the 2 GPa change will be discussed further below, the 4 GPa change is most likely associated with the $T_\textrm{N} (P)$ line crossing the solidification temperature line ($T_\textrm{s}(P)$) for the 1:1 mixture of $n$-pentane:iso-pentane pressure medium near 4\,GPa.\cite{Torikachvili2015}  

The residual resistivity ratio of CrAuTe$_4$ also changes with pressure and manifests a very clear, non-monotonic behavior (Fig.\,\ref{Pressure_anal}\,(b)). The broad, local maxima between 1.7 and 2.1 GPa is followed by a shallow drop at higher pressures.

Figure\,\ref{Pressure}\,(b) shows the MR ratio (defined as $(\rho(H)-\rho(H=0))/\rho(H=0)$)for $T$\,=\,1.8\,K measured at 14 different pressures. To be consistent, we always applied the current along the $b$-axis and magnetic field along the $c$-axis. MR shows quadratic behavior at low pressure, and near-linear behavior is detected at high pressure. To be more quantitative, we fitted the MR data with $c H^n$, where c is a constant, H is the field, and n is an exponent.The values of n are plotted in Fig.\,\ref{Pressure_anal}\,(c). n is around 2 at low pressure, but starts to drop from 2 GPa. Finally, $n \sim 1$ around 5.22\,GPa. Conventionally MR is proportional  to $H^2$, but it sometimes shows linear MR when there is scattering and magnetic breakdown.\cite{Pippard1989} More recently, A. A. Abrikosov suggested that linear energy dispersion can induce linear MR.\cite{Abrikosov1998} Although sudden increases in scattering (with pressure) are possible, magnetic breakdown happens over limited high field regimes. However, electronic transition is possible with introducing linear dispersion. It is possible, then, that the anomaly in MR exponent is a result of the electronic phase transition under pressure.

\begin{figure}
	\includegraphics[scale=1]{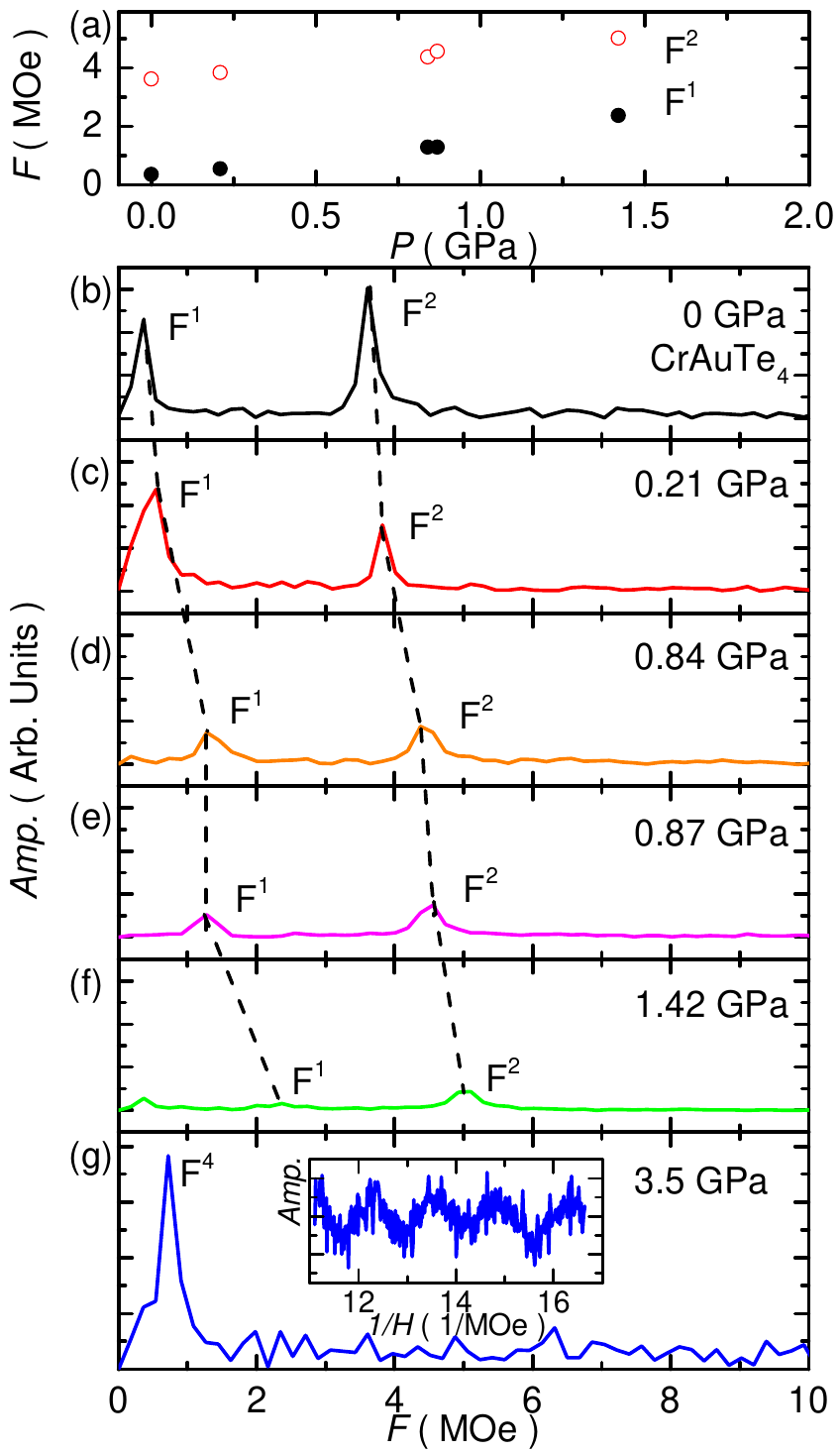}%
	\caption{(color online) FFT results of SdH oscillation from 0\,GPa to 1.42\,GPa, and 3.5\,GPa (a) overall trend of frequencies change as a function of pressure. (b) 0\,GPa, (c) 0.21\,GPa, (d) 0.84\,GPa, (e) 0.87\,GPa, (f) 1.42\,GPa, (g) 3.5\,GPa. Inset of (g) is the periodic oscillation as a function of 1/H after subtracting background. 
		\label{Frequency}}
\end{figure}

 Figure \ref{Pressure_anal}\,(c) also shows the MR at 90\,kOe. The MR is 1.85 at ambient pressure, with a local maximum at 0.87\,GPa of 3.45, and then drops down to 0.57 at 5.22\,GPa. It noticeably starts to drop around 2\,GPa, and, as expected, at higher pressures shows a similar pressure dependence as the exponent, n. Given that the pressure dependence of $T_\textrm{N}$, RRR, the MR as well as the MR power-law exponent all show changes in behavior just below 2\,GPa, it is likely that there is a pressure induced transition near this pressure.      
 
The pressure dependence of the SdH oscillations was tracked as well. As pressure increases, both frequencies move to higher values (Fig.\,\ref{Frequency}). This implies that the respective Fermi surfaces areas in k-space increase in the ab plane.The pressure dependences of the changing frequencies are presented in Fig.\,\ref{Frequency}\,(a). Both frequencies, $F^1$ and $F^2$, show almost linear change. The pressure derivatives of extremal cross sections which is defined as $dlnF/dp$ are calculated for each frequencies; $dlnF_{1}/dp = 1.31$ GPa$^{-1}$ and $dlnF_{2}/dp = 0.23$ GPa$^{-1}$. The pressure derivative of $F_2$ is comparable to previous results on Bi and Cd\cite{Schirber1970,Budko1984}, but $F_1$ is at least 2 times greater. 
 
 Not only frequencies but also amplitudes are changed as pressure is increased. As $F^1$ and $F^2$ are smeared out, no more frequencies are detectable in the PCC for $P\,>$\,1.42\,GPa.. This may suggest that the local curvature near extremal cross-sections of the Fermi surface is changing as we apply pressure. In addition, we start to seeing a new low frequency peak from 2.08\,GPa, and finally pronounced low frequency peak, $F^{4} = 0.71$\,MOe, is observed at 3.5\,GPa. Low frequency peaks can sometimes be due to an artifact during the data processing. However, $F^4$ has large amplitude and, as shown in the inset of Fig.\,\ref{Frequency}\,(g), it can even be clearly seen in the raw data.  
 
Based on the present data, it is apparent that the antiferromagnetism in CrAuTe$_4$ is not particularly "fragile" and much higher pressures would be needed to suppress it.  On the other hand, there is a pressure induced phase transition near 2\,GPa.  We cannot identify the nature of this transition; it could be a pressure induced change in the magnetic ordering wave vector, it could be a structural phase transition, or it could be a Lifshitz transtion.  
  
\section{Summary}

In summary, we studied physical properties of high quality single crystal, CrAuTe$_4$. The temperature dependent magnetization has been measured, and we observed the anisotropic susceptibility consistent with the spins aligned antiferromagnetically along the $c$-axis below 255\,K. The temperature dependent resistivity also showed a clear, loss-of-spin-disorder-scattering feature at the antiferromagnetic transition temperature. The 2 K MR showed anisotropy with factor of 2, SdH oscillations are observed and analyzed. The band structure was measured by ARPES below and above the Neel temperature; band folding due to the antiferromagnetic transition is detected. To test for possible fragile magnetism, the pressure dependence of the temperature dependent electrical resistivity was measured for $p$\,$<$\,5.25\,GPa. $T_\textrm{N}$ was suppressed to 236\,K by 5.22\,GPa.  Suggesting a possible pressure induced phase transition that affects the electronic properties. 

\begin{acknowledgements}
Authors would like to thank M. C. Nguyen for very helpful discussions. Research was supported by the U.S. Department of Energy, Office of Basic Energy Sciences, Division of Materials Sciences and Engineering. Ames Laboratory is operated for the U.S. Department of Energy by the Iowa State University under Contract No. DE-AC02-07CH11358. Na Hyun Jo is supported by the Gordon and Betty Moore Foundation EPiQS Initiative (Grant No. GBMF4411).
\end{acknowledgements}

\clearpage

\bibliographystyle{apsrev4-1}
%
\end{document}